**Predicting Human Touch Sensitivity to Single Atom Substitutions in Surface Monolayers**

Abigail Nolin[1], Amanda Licht[1], Kelly Pierson[1], Laure V. Kayser[1,2], Charles Dhong[1,3]*

[1]Department of Materials Science & Engineering, University of Delaware, Newark, DE 19716, USA

[2]Department of Chemistry and Biochemistry, University of Delaware, Newark, DE 19716, USA

[3]Department of Biomedical Engineering, University of Delaware, Newark, DE 19716, USA

*Author to whom correspondence should be addressed: cdhong@udel.edu

**Abstract**

The mechanical stimuli generated as a finger interrogates the physical and chemical features of an object forms the basis of fine touch. Current approaches to study or control touch primarily focuses on physical features, but the chemical aspects of fine touch may be harnessed to create richer tactile interfaces and reveal fundamental aspects of tactile perception. To connect tactile perception with molecular structure, we systematically varied silane-derived monolayers deposited onto imperceptibly smooth surfaces and made predictions of human tactile sensitivity via friction and cross-correlation analysis. We predicted, and demonstrated, that humans can distinguish between two isosteric silanes which differ only by a single nitrogen-for-carbon substitution. The mechanism of tactile contrast originates from a difference in monolayer ordering which was replicated in two alkylsilanes with a three-carbon difference in length. This approach may be generalizable to other materials systems and lead to new tactile sensations derived from materials chemistry.

**Main text**

Fine touch is used to discriminate between objects, to determine quality, and is a critical source of information for people with low vision or blindness. Tactile stimuli are known to be generated when a finger runs across physical features, such as bumps or other mechanical textures.(*1–3*) While physical features are used to design surfaces for touch, including active control via haptic technologies(*3–8*), objects do not only contain distinctive physical features. They also possess attributes originating from surface chemistry and molecular structure. The limits of generating tactile stimuli solely from materials chemistry, in the absence of physical features or mechanical actuation, is unknown. To connect surface chemistry and tactile perception, mediated by friction and adhesion phenomena, we studied and built predictions of what types of chemical structures are perceivable by human subjects on physically smooth surfaces. Chemical control over tactile sensations, and the ability to rapidly screen new tactile materials, would rapidly expand the breadth of accessible synthetic tactile sensations for haptic technologies and provide basic knowledge into tactile perception.(*9*)

To connect molecular structure with tactile sensations, we used a model system of silane-derived monolayers on silicon wafers. Silanes form monolayers on surfaces with comparative ease.(*10*) These monolayers and the underlying silicon wafers are smoother than the human limits of tactile perception of surface roughness (~10 nm) (*3*), minimizing macroscopic physical contributions to tactile sensations. The diversity of silanes offers a practical route to systematically decouple factors influencing adhesion and friction through materials selection. In short chain alkylsilanes, increasing the chain length greater than *N*=6 (where *N* refers to the number of carbons or nitrogen atoms in the functional group of the silane) results in a decrease in friction by atomic force microscopy(*11*) despite the fact that the surface energy, an estimate of average friction force, of simple alkylsilanes are almost identical. This abrupt decrease in friction for alkyl chains longer than *N*=6 originates from a phase transition from a disordered to ordered (solid-like) monolayer due to increased van der Waals interactions, which reduces the number of modes for energy dissipation in the monolayer.(*11*) Functionalized silanes, such as aminosilanes, contain amine groups which further increase monolayer ordering through hydrogen bonding as compared to alkylsilanes. In addition, the polarity of amine groups increases surface energy.(*12*) Although the friction of silanes is often studied at the microscale, these results can only be used qualitatively to understand soft, mesoscale systems like the human finger because microscale friction replicates the interaction between a single asperity and a surface.(*13*)

To rapidly screen multiple silane coatings prior to human testing, we sought to predict the human ability to distinguish materials based on the expected mechanical stimuli present during a touch event. Mechanical stimuli is created by the friction generated by a finger sliding across an object, which is ultimately transduced by afferent nerves and give rise to a tactile percept.(*14–17*) Friction depends on force and sliding velocities, both of which are variable when humans freely explore objects. Therefore, developing a general relationship between touch and chemical structure is difficult because friction is not a material property.(*2*, *18–20*) One approach to address the variability of friction is to categorize surfaces by their average friction force, i.e., a constant coefficient of friction, under a subset of exploration conditions. However, the use of an average friction force in fine touch leads to an ineffective framework: it suggests that two dissimilar surfaces should feel the same by pressing harder on the surface with a lower friction coefficient. Others have incorporated friction as a dynamic quantity, including within the context of large physical features (~100 μm—mm)(*1*, *21*, *22*), but these studies are either *post hoc* or are not generalizable to tactile stimulus derived from chemical origins. Separate from challenges in the mechanics of predicting mesoscale friction, it is currently unresolved what portions of the friction traces are critical to the human ability to distinguish surfaces.(*3*, *9*, *14*, *23*, *24*) Here, we posit that subjects distinguish surfaces by the distinctive oscillations in friction forces which originate from stick-slip phenomena: deviations from steady sliding due to the transient trapping and energetic release of adhesion from asperities.(*18*, *23*, *25*, *26*) (see **S2** in **Supplemental Materials** for brief explanation) To address the lack of predictive models on human performance, we performed mesoscale friction measurements at multiple conditions similar to human exploration and interpreted our results based on the total differences in stick-slip friction present. Our approach treats friction as a dynamic quantity, addresses the human variability of free exploration, and is independent of unresolved correlations between friction traces and human tactile perception.

The selection of silanes, and characterization, in **Table 1** encompasses a range of alkyl and aminosilanes around the *N*=6 transition. Silane characterization with contact angle hysteresis and atomic force microscopy shows low surface roughness, similar surface energy amongst alkylsilanes, and low contact angle hysteresis. (scan area of 10 μm × 10 μm, $R_a$ profile along diagonal ~14 μm, see **S3** in **Supplementary Materials**) Therefore, the model system has minimized physical contributions to friction, isolated the role of average adhesion versus differences in friction oscillations due to stick-slip phenomena, and has low physical and chemical heterogeneity. Terminal and secondary amines have increased intermolecular forces from hydrogen bonding, but terminal amines have a lower water contact angle than secondary amines.

Therefore, **C4-APTMS** and **APTMS,** in conjunction with alkylsilanes, help clarify the role of amino functional groups in influencing friction through surface energy and monolayer ordering.

**Table 1. Silane precursors and properties**

| Silane precursor | Surface structure | Water contact angle (°) | Roughness, $R_a$ (pm) |
|---|---|---|---|
| (3-aminopropyl)trimethoxysilane (**APTMS**) | 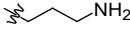 | (31.1-48.2) ± 2.3 | 222 |
| n-methylaminopropyltimethoxysilane (**C1-APTMS**) | 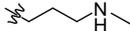 | (24.5-37.9) ± 1.8 | 341 |
| n-butylaminopropyltimethoxysilane (**C4-APTMS**) | 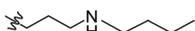 | (56.8-72.7) ± 2.2 | 209 |
| n-butyltrichlorosilane (**C4**) | 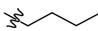 | (94.7-104.0) ± 2.9 | 455 |
| n-pentyltrichlorosilane (**C5**) | 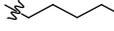 | (99.8-106.9) ± 1.5 | 410 |
| n-hexyltrichlorosilane (**C6**) | 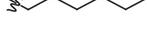 | (97.8-109.7) ± 0.7 | 471 |
| n-heptyltrichlorosilane (**C7**) | 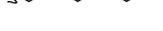 | (98.4-103.5) ± 0.9 | 662 |
| n-octyltrichlorosilane (**C8**) | 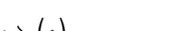 | (97.2-103.8) ± 2.7 | 321 |
| n-octadecyltrichlorosilane (**C18**) | 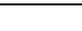 | (100.8-106.9) ± 0.6 | 337 |

To collect stick-slip friction measurements of silanes for predictions of human performance, a custom mechanical apparatus setup was used to measure the friction of an elastic, but not sticky, mock finger at pressures ($M = 0 – 100$ g, added to the deadweight of the mock finger) and velocities ($v = 5 – 45$ mm/s) consistent with human exploration and sampling rates sufficient to capture signals important for the mechanoreceptors important in fine touch (FA-I mechanoreceptors operate up to ~40 Hz). (*16*, *18*) (**Figure 1A**, and section **S1** in **Supplementary Materials**). Friction traces generated by a mock finger sliding across **C5** and **C4-APTES** are shown in **Figure 1B**. For **C5** at $v = 25$ mm/s and an applied mass of $M = 0$ g (i.e., a mock finger with no added mass), the friction oscillations appear relatively even, but a distinct stiction spike appears when the finger is slid with a higher applied mass of $M = 25$ g. A distinct stiction spike is also visible on **C4-APTES** under the same conditions. These stiction spikes are conspicuous and have been suggested as one route to explain tactile discriminability in fine touch.(*14*) However, in the surfaces studied here, the disappearance and reappearance of stiction spikes at different conditions on the same surface suggests that stiction spikes are incomplete indicators of the ability for humans to identify surfaces.

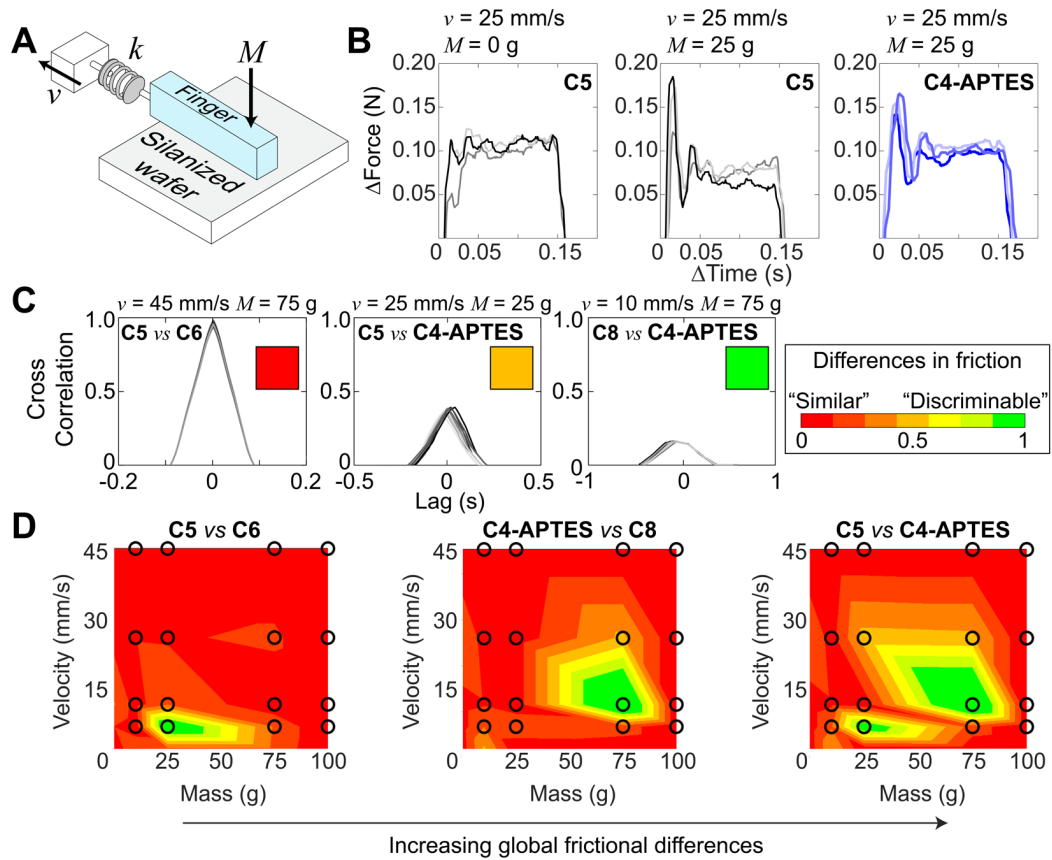

**Figure 1. Friction measurements and cross-correlation analysis.** (A) Schematic of mechanical testing apparatus, where an elastic mock finger is slid across silanized silicon wafers at a sliding velocity, $v$, and an applied mass, in addition to the finger, $M$. Force transduced by sensor with spring constant $k \gg$ finger elasticity. (B) Representative friction traces from the same surface (left, middle) at different conditions, and different surfaces (middle, right) at the same conditions. (C) Cross-correlation of friction traces from two surfaces at different conditions, quantifying the similarity (or differences) in friction traces. Similar friction traces are likely difficult to discriminate for subjects and noted in a red box, whereas distinct friction traces are likely easier to distinguish are noted with a green box. (D) Discriminability matrices summarizing differences in friction traces between two silanes across experimental conditions. Differences are quantifying the skew in the cross-correlation curve. Black circles represent conditions at which experiments were performed, heat map created by 2D interpolation. Friction was measured in triplicate, then repeated on three fresh spots per velocity, mass, and silane (144 traces per silane).

To synthesize the experimental space of masses, velocities, and silanes into a prediction of tactile performance, we quantified the total similarities or differences between friction traces through cross-correlation.(*18*, *23*) (Methods in **S5** of **Supplemental Materials**) Humans use friction as a tactile cue(*2*, *3*, *14*, *23*) and two surfaces which generate similar friction forces under most experimental conditions are probably more difficult for subjects to distinguish, whereas two surfaces which generate distinct friction forces are probably easier for subjects to distinguish. The degree of similarity present between the friction forces of two silanes was quantified by cross-correlating friction traces obtained from each silane at a given applied mass and velocity. A representative cross-correlation with a high (red), medium (orange), and low (green) similarity is shown in **Figure 1C**. A large and symmetric, i.e. triangular, cross-correlation trace indicates that the friction traces are similar. The amplitude of the cross-correlation was normalized to account for different sliding velocities and force magnitudes, and then the symmetry of the cross-correlation trace was parameterized into a single value by the skew. This process condensed multiple friction traces from two silanes into a single value on the scale of "similar" to "discriminable" (red-to-green scale of skew used in **Figure 1C**, **D**, skew normalized for convenience, distribution in **Figure S4** of Supplemental Materials).

Experimental conditions which led to similar friction traces (large cross-correlation) are shown in red and conditions with distinct friction traces are shown in green (small cross-correlation). (see **Figure 1D**, each matrix represents 144 × 2 friction traces) In contrast to a standard analysis of friction coefficients, these matrices identifies both pairs of silanes and experimental conditions under which two different silanes are likely to be distinguishable to human subjects. At $M = 25$ g and $v = 5$ mm/s, **C5** and **C6** appear to yield distinct friction forces from the green region in **Figure 1D**. However, under most other experimental conditions, the two surfaces are red and likely to be difficult for human subjects to distinguish. In contrast, the **C5** vs **C4-APTES** matrix has regions of green under several applied masses and sliding velocities, indicating conditions that the two surfaces are likely easier to distinguish under a wide variety of exploration conditions. Notably, one condition where **C5** vs and **C4-APTES** is not discriminable is when both presented a stiction spike highlighted in **Figure 1B** ($M = 25$ g, $v = 25$ mm/s), reducing the role of stiction spikes in predicting tactile performance.

The discriminability matrices help address a drawback in predicting human behavior from a narrow experimental space. **C5** vs **C4-APTES** might have been prematurely excluded from the basis of a few measurements which indicate low discriminability (red) at $v = 45$ mm/s, but upon testing under other experimental conditions, green regions indicate that the two silanes might be good candidates for human testing. Another material pair, **C4-APTES** vs **C8,** only differ by a single site substitution (isosteric silanes), but the presence of several green regions indicate that the two surfaces might also be discriminable by human subjects. If human performance is related to the number of green areas in the discriminability matrices, mechanical testing indicates that **C5** and **C4-APTES** is more likely to be distinguished (or discriminable) by human subjects than either **C5** vs **C6** or **C4-APTES** vs **C8**,

To verify if humans could tell surfaces apart by touch, we performed a three-alternative forced choice test ("odd-man-out" test, see power analysis in **S6** of **Supplemental Materials**). Subjects were presented three surfaces: two surfaces coated with a silane and one surface with another silane. Subjects were asked to identify the "odd-man-out", or the one surface that was unlike the other two in five sequential trials. The location and silane used as the single alternative was randomized across trials. Subjects could freely explore surfaces, were not restricted in how they touched surfaces and were not given visual or auditory blinds. (Incidentally, all surfaces were visually indistinguishable) Our free exploration conditions closely mimicked how subjects interact with objects in everyday scenarios, improving ecological validity and the generalizability of our findings.(*27*) Finally, subjects were untrained, and it was highly unlikely that subjects had felt any of these surfaces before. Results from human testing are shown in **Figure 2A**.

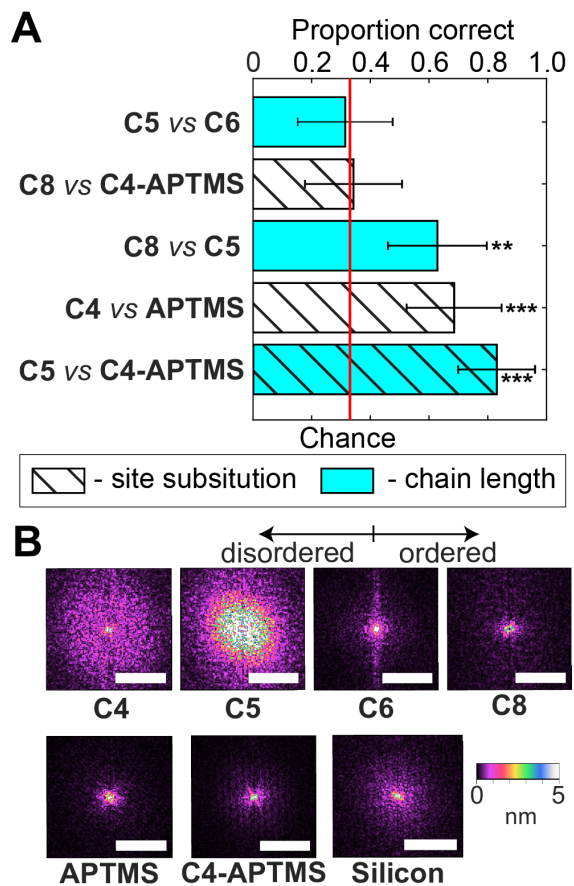

**Figure 2. Human psychophysical testing.** (A) Results of "odd-man-out" (three-alternative forced choice) psychophysical testing with different pairs of silanes. Error bars indicate 95% confidence intervals. Asterisks represent significance level. Each test had thirty-five trials with $n=7$ subjects. (B) Monolayer order versus disorder by Fourier transforms of surface profiles obtained from atomic force microscopy. Silicon represents bare wafer as a control. Scale bar = 10 μm$^{-1}$.

Human subjects successfully discriminated between three pairs of surfaces, shown in **Figure 2A**. Subjects could discriminate between **C8** vs **C5** (accuracy = 62.8%, $t = 3.56$, Cohen's $d = 0.60$, $p < 0.005$)—surfaces which differ by a length of three carbons. The underlying mechanism, or tactile contrast, between **C8** and **C5** is likely to due to the increased order from the surface monolayer from the longer alkyl chain, which decreases friction.(*11*) This transition between **C5** and **C8** is evident in the Fourier transform of surface profiles obtained by AFM (**Figure 2B**, power spectral densities shown in **Figure S2B-C**). **C5** exhibits a more globular structure than **C8**, evidenced by the larger, diffuse FFT (fast Fourier transform). In contrast, the more ordered, solid-like **C8** better recapitulates the underlying silicon. At the $N=6$ transition, the FFT of **C6** appears to demonstrate an incipient phase change by the increased globular structure compared to the silicon. However, this incipient phase change appears too subtle for human subjects and **C5** vs **C6** was not discriminable in human testing (average accuracy = 31.4%, at chance). Together, results from human testing on the incomplete phase change between **C5** vs **C6** and the more realized phase change between **C5** vs **C8** demonstrate that monolayer ordering is a new route to create tactile contrast between surfaces, but only if the phase transition occurs to a sufficient degree.

An alternative method of using chain length to increase monolayer order is to substitute a carbon for an amine group to provide hydrogen bonding. Results from human testing showed that subjects were able to distinguish between **C4** vs **APTMS** (accuracy = 68.6%, $t$ = 4.43, Cohen's $d$ = 0.75, $p < 0.0001$), even though both silanes are isosteric. The source of tactile contrast is evident in the FFT, where **C4** is globular and disordered, whereas **APTMS** creates an ordered and periodic structure, even below the $N=6$ transition. However, when comparing **C8** vs **C4-APTMS**, which are long ($N = 8$) alkyl and aminosilanes, subjects were unable to discriminate between **C8** vs **C4-APTMS** (accuracy = 34.2%, at chance). The single amine substitution only created tactile contrast when the additional intermolecular forces induced a phase change, which no longer occurs when silanes are sufficiently long. Therefore, single site substitutions alone are not always noticeable to humans but become noticeable if they surpass a critical threshold in intermolecular forces to induce a phase change.

Our results suggest that tactile contrast, i.e., human performance, is highest when maximizing differences in order and disorder between two surfaces. By comparing **C5** vs **C4-APTMS**, we combined effects from chain length and phase changes induced by amine substitution, as confirmed by FFT. As expected, subjects had the highest accuracy out of all comparisons when discriminating **C5** vs **C4-APTMS** (accuracy = 82.3%, $t$ = 7.66, Cohen's $d$ = 1.30, $p < 1 \times 10^{-8}$). Differences in order and disorder were more consistent in explaining human performances than surface energy or roughness. Demonstrating the inconsistencies of surface energy, both **C4** vs **APTMS** and **C8** vs **C4-APTMS** exhibit moderate differences in contact angle (>35° see **Table 1**), however only **C4** vs **APTMS** was discriminable by human subjects. Furthermore, **C8** vs **C5** was also discriminable, but have nearly identical surface energies.

Prior to insights from human testing, the discriminability matrix based on skew in **Figure 1** was a convenient starting point for predicting human performance on different materials via friction. To better explain human performance, results from human testing were fitted to several predictors based on the friction traces with a stepwise linear regression. To perform a regression, predictors must be identified. However, there was little previous guidance about which aspects of friction traces are most important for human tactile discriminability. We formed predictors by summing a parametric value across all masses and velocities for a given pair of silanes from our cross-correlation analysis. (These parametric values were the average, variance, skew, and kurtosis, of the cross-correlation.) We also considered "zones of discriminability" as predictors. Zones of discriminability count the number of green regions, i.e. high discriminability, in **Figure 1**, which are not apparent when taking an average value of discriminability. This was quantified by counting the number of experimental conditions in the top quartile of each parametric category (e.g., skew, variance) across every surface. **Figure 3** shows the results of the linear regression model.

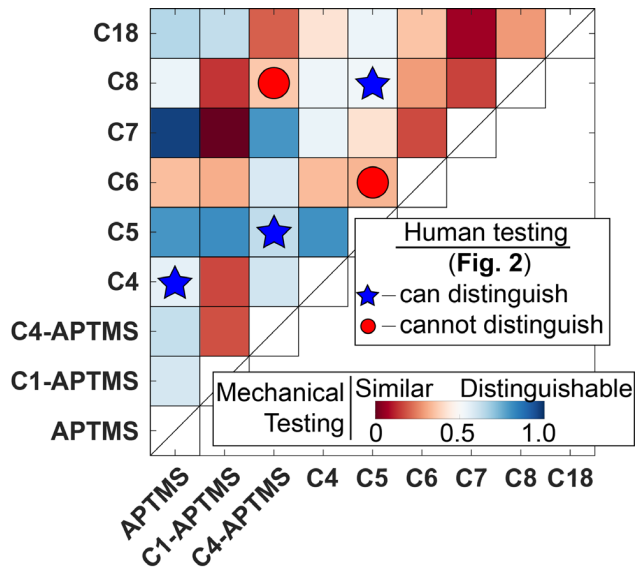

**Figure 3**. **Modeling human responses from mechanical testing.** Linear regression fit of human performance (**Fig. 2**) with mechanical testing data across all combinations of silanes. Stars and circles indicate pairs of silanes which human subjects could and could not distinguish, respectively.

The linear model, shown in **Eq. 1.,** yielded a correct predictive response on each of the five different silanes tested. ($t$(variance) = -6.28 and $p < 0.01$, $t$($n_{Q4}$ skew)) = 11.83 and $p < 0.01$, $F = 65.7$, $p < 0.001$, $r^2 = 0.71$ for skew alone, $r^2 = 0.96$ with both terms).

$$Distinguishability = 2.22 - 2.06 * \Sigma\ variance + 0.20 * \Sigma\ (n_{skew} \geq Q_4) \qquad (1)$$

Here, "distinguishability" is a scale from 0 to 1 which denotes pairs of materials and is not the same scale as the earlier use of "discriminability" which was a comparison of friction traces. The total variance from mechanical testing was negatively correlated with human performance which suggests that a wide distribution of different friction forces was not helpful for discriminating surfaces. Rather, a smaller (but presumably distinct) distribution of amplitudes in the oscillations of friction contributes to the ability of subjects to discriminable samples. The importance of zones of discriminability based on skew offer a clearer mechanistic origin for the human ability to discriminate between silanes: unlike averaging the entire experimental space, the zones of discriminability are traceable to discrete experimental conditions. Skew represents asymmetry between the number of large amplitudes versus small amplitudes, relative to the mean amplitude. Therefore, to create better tactile contrast in materials, the model suggests materials where one surface generates large amplitude oscillations in friction forces and the other surface generates low amplitude oscillations at a given applied mass and sliding velocity. Another route for improved tactile contrast is that optimizing on a few, but distinct, experimental conditions is more important than creating surfaces with moderate differences across multiple conditions. In the material system here, we generated high skew surfaces by combining the effects of a short and long chain length with the effects of hydrogen bonding to create an abrupt and complete phase transition. Even in relatively simple alkylsilanes, another potential route of tactile contrast, the odd-even effect of silanes(*28*), is suggested by the higher model prediction on **C7** than **C8** compared to **APTMS**.

Here, we used mechanical testing and human psychophysics to design tactile surfaces based on molecular phenomena originating from physically smooth surfaces made with silane-derived monolayers. We predicted and confirmed that humans are sensitive to changes as subtle as a single atom substitution of a carbon-to-nitrogen between isosteric alkyl and aminosilanes, or alkyklsilanes with a three carbon difference in chain length, which identified monolayer ordering as a new route to create tactile sensations. Discovering that humans are sensitive to molecular effects as small as a single site substitution opens the door to chemical and materials approaches for the rationale design of new tactile sensations. Furthermore, we developed a method to establish structure-property relationships for tactile materials by connecting materials phenomena to human tactile performance via mesoscale friction. This methodology can be immediately applied to repurpose existing material platforms, such as stimuli-responsive polymers and liquid crystal elastomers, to discover new tactile sensations. This will lead to richer and more accurate tactile sensations for human-machine interfaces, higher quality tactile graphics for the blind, enable no-power tactile diagnostics of cognitive function, and modernize tools to investigate fundamental aspects of touch by providing sufficiently precise tactile interfaces to study sensory integration and perception.

**Acknowledgements**

We acknowledge the Keck microscopy core at the University of Delaware for AFM assistance and the Biostatistics, Epidemiology and Research Design core for discussions on statistical analysis.

**Participants.** This study was conducted and approved by the Institutional Review Board of the University of Delaware (Project #1484385-2). Data were collected from a total of 14 healthy volunteers between the ages of 18 and 40. **Author Contributions**. A.N., L.K., and C.D. wrote manuscript, designed experiments, and collected data. A.L. and K.P. designed experiments and collected data. **Funding.** Support from the University of Delaware.

# Supplemental Materials

for

# Predicting Human Touch Sensitivity to Single Atom Substitutions in Surface Monolayers

Abigail Nolin[1], Amanda Licht[1], Kelly Pierson[1], Laure V. Kayser[1,2], Charles Dhong[1,3]*

[1]*Department of Materials Science & Engineering, University of Delaware, Newark, DE 19716, USA*
[2]*Department of Chemistry and Biochemistry, University of Delaware, Newark, DE 19716, USA*
[3]*Department of Biomedical Engineering, University of Delaware, Newark, DE 19716, USA*
*Author to whom correspondence should be addressed: cdhong@udel.edu

## S1. Methods and experimental rationale

*Surface preparation.* Silanes were purchased from Gelest (95-100%) and used as is. Silanes consist of a reactive headgroup to anchor to a surface and a variable tail. The variable tail consists of alkyl chains or other functional groups. Silicon wafers (University Wafers) were cleaned with acetone, IPA, DI and then dried. Surfaces were then exposed to oxygen plasma (Glow Plasma System, Glow Research) for 30 seconds. Wafers were immediately transferred to vacuum desiccators containing 50 μL of silane on a glass coverslip. Desiccators were then evacuated and held under static vacuum for 4h, except for OTS which is known to require both heat (60 °C) and >24h exposure.(*1*) Separate desiccators were used to prevent potential cross-contamination. Plasma treatment and silane treatments were confirmed with water contact angle.

*Mock finger preparation.* To recreate these macroscopic sliding friction conditions, we mimicked the finger with a rectangular PDMS slab with dimensions of 1 cm × 1 cm × 5 cm. A rectangular geometry, unlike a hemispherical or cylindrical geometry, retains a consistent nominal contact area even under different loading conditions.(*2*) We believed this load-independence better mimicked the motion of a human finger sliding across a surface than a spherical indenter: Although the fingertip appears rounded, subjects tended to leave residue with a largely consistent width. This is better mimicked by a rectangular finger which maintains a constant width. Furthermore, the expected contact area between a finger and object approaches the majority of the contact area on a finger quickly, even at low loads.(*3*) Mock fingers were formed by casting polydimethylsiloxane (PDMS) into a rectangular mold and curing at 60° C for 24 h. Inside the mold, an acrylic, 3D-printed bone provided mechanical rigidity and replicated the mechanical stiffness by the distal phalange within a human finger. The elastic modulus of the PDMS was controlled by the ratio of base to crosslinker (30:1) to achieve 100 kPa, similar to the effective modulus of human fingertips.(*4*) As the finger is a soft, but not sticky object, we also treated the surface of the PDMS with a UV/O treatment, a mild but long-lasting surface coating which results in a hydrophilic surface coating with a similar water contact angle to that of human skin.(*5*)

*Human sliding velocity and applied mass.* For our mechanical tests, we used conditions largely consistent with other studies on fine touch. We asked subjects to slide their finger across samples while tracking their motion with a small (~1 mm, <5 g) motion capture device at 240 Hz and ~0.02 mm accuracy (Polhemus Liberty Motion Tracking). (see **Figure S1**) Exploration velocities ranged from 13.5-42.8 mm/s. For applied mass, subjects were asked to slide their finger across a wafer on a scale. Applied masses ranged from 9.3 g to 93.8 g. Subjects indicated they were able to press harder but did not typically exceed 100 g for discrimination tasks.

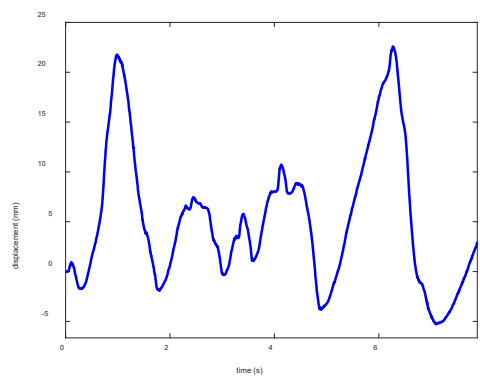

**Figure S1. Motion tracking of subject finger**. Subject motion tracking data used to determine exploration velocity conditions in mechanical tests.

## S2. Stick-slip Phenomenon

Stick-slip phenomena arise when portions of a sliding finger are mechanically stuck on a barrier (stick) due to adhesion or surface defects. As the bulk continuously builds up force, this stuck portion overcomes and overshoots this barrier (slip). The elasticity of the sliding object, i.e., the finger, permits for a heterogeneous distribution of global and local stick-slip events across the finger.(*6*) Stick-slip is common in most experimental systems and responsible for the sound made by squeaky wheels, violins, and crickets, and the tremors of earthquakes.(*2*) For many applications, the oscillations due to stick-slip friction are averaged into a bulk coefficient of friction.

## S3. Silane Characterization

*Atomic Force Microscopy*. Surface profiles were obtained through a tapping mode in atomic force microscopy (Veeco Dimension 3100 Scanning Probe Microscope, Veeco and Gwyddion SPM analysis software) over a scan area of 10 μm × 10 μm at a scan rate of 2 μm/s and drive frequency of 316 KHz. Example of results are in **Figure S2**A. Surfaces over a relatively large area show physiosorbed aggregates, more prevalent in **C4**.(*7*) The power spectral densities (PSD)(*8*) of the alkylsilanes in **Figure S2B** and aminosilanes in **Figure S2C** indicate a more disordered surface in alkylsilanes below a length of C8. The PSD of C8 and C18 appear more like the bare silicon control. C6 appears to show less disorder than C7, potentially due to less efficient chain packing off odd-length chains.(*9*)

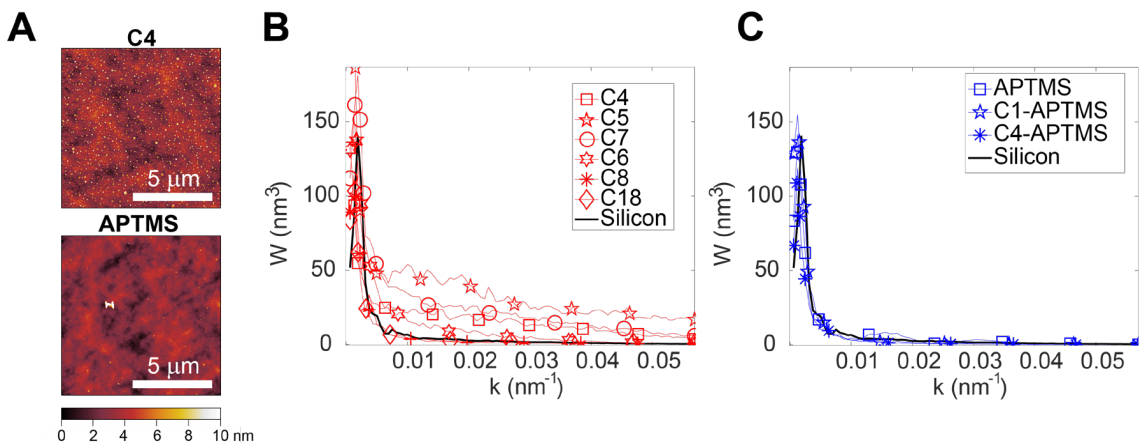

**Figure S2. Surface profiles by atomic force microscopy**. (A) Surface profiles of C4 and APTMS. Scan area of 10 μm × 10 μm. (B) Power spectral density (PSD) of surfaces of alkylsilanes to characterize roughness. (C) PSD of aminosilanes.

*Contact Angle Hysteresis*. Advancing and receding water contact angles were measured for each silane with a goniometer (DSA14 Drop Shape Analysis System, Kruss). To measure the advancing angle, an ~4μl DI water drop was first dispensed onto the silane surface. Small increments of water were added with a syringe attached to micromanipulator until the drop moved. Once the drop visibly spread, an image was captured (see representative images in **Figure S3**). To measure the receding angle, water was slowly removed from the same drop until the drop visibly retracted, and an image was captured. Five drops were measured for each silane surface. The contact angle of each image was analyzed using an automatic circle fit (ImageJ). Hysteresis was calculated by taking the difference of each advancing and receding angle pair. The average receding angle, average advancing angle, and standard deviation of the hysteresis is reported in **Table 1**.

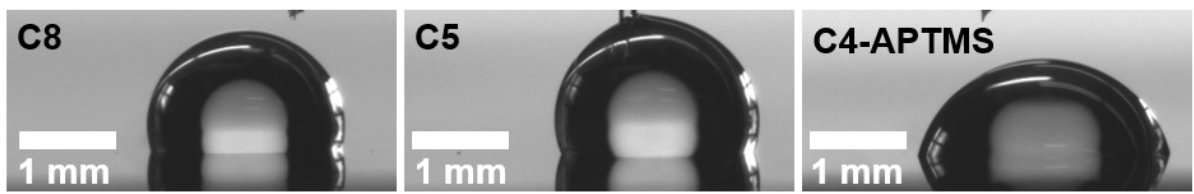

**Figure S3**. **Contact Angle Hysteresis**. Representative images of advancing angles from contact angle measurements. The needle tip is visible in the middle panel and on the periphery of the other panels.

## S4. Mechanical Apparatus

We sought to mimic the macroscopic friction generated by a subject as they slide their finger across a surface. Compared to the typical laboratory techniques for measuring friction, such as an AFM, the friction generated by a human is a coarse method of characterizing surfaces. That is, for humans to make judgements about two surfaces based on differences in friction on each surface, these differences in friction must be significant compared to typical friction tests. Any differences in friction must be robust enough to be consistent across multiple microscopic variations, such as dust, precise contact area, and device alignment. The PDMS mock finger was loaded with a desired additional mass, $M$, and then brought into contact with the surface with a micromanipulator until it reached a contact length of 1 cm from the side, visually. This resulted in a nominal contact area of 1 cm × 1 cm. The finger was then slid for a distance of 4 mm with a magnetic motorized stage (V-508 PIMag Precision Linear stage, Physikinstrumente) at a constant velocity, $v$, (verified with an internal laser displacement sensor) across the surface. This sliding was repeated 4 times. Simultaneously, the friction force on the finger was measured with a Futek 250 g LSB ($k$ = 13.9 kN m$^{-1}$, peak-to-peak noise of 0.1 mN) sampling at 550 Hz (Keithley 7510 DMM). We chose our sliding distance because human subjects are unlikely to slide at a consistent velocity for long distances. Of the four distinct slides, data from the first slide was always discarded because it may have included extraneous pressure or aging as the test was being setup. In total, we obtained three slides from each of three fresh spots for each condition (i.e., for a given mass, velocity, and surface functionalization).

## S5. Cross-correlation and skew analysis.

The cross-correlation is normalized by the absolute magnitude of both input vectors is calculated by:

$$Cross - \widehat{Correlation} = \frac{\sum\left((a(t) - \hat{a}) * (b(t) - \hat{b})\right)}{\sqrt{\sum(a(t) - \hat{a})^2} \sqrt{\sum(b(t) - \hat{b})^2}} \quad (S1)$$

Where *a* and *b* are the time-series friction measurements and *â* represents the mean value of the vector. The skew of the cross-correlation is normalized to have a maximum value of 1 based on all measurements simply for convenience in plotting and to highlight the concept that descriptive parameters (variance, kurtosis, etc.) of the cross-correlation are comparable through discriminability matrices. A distribution of skews is shown in **Figure S4**. On the left is the skew from each cross-correlation at each mass and velocity per comparison of pairs of silanes. On the right is the skew summed from all masses and velocities per comparison of pairs of silanes. The scale factor for skew used in the discriminability matrices is 3.17, which is evident from normalizing the maximum skew encountered in this experiment, which was 0.315. Normally, the sign of the skew indicates whether the skew is left or right. However, we took the absolute value of skew at under each experimental condition, so a single silane with large amounts of left or right skew would result in a surface with overall high skew and that the left and right skew regions do not cancel each other out.

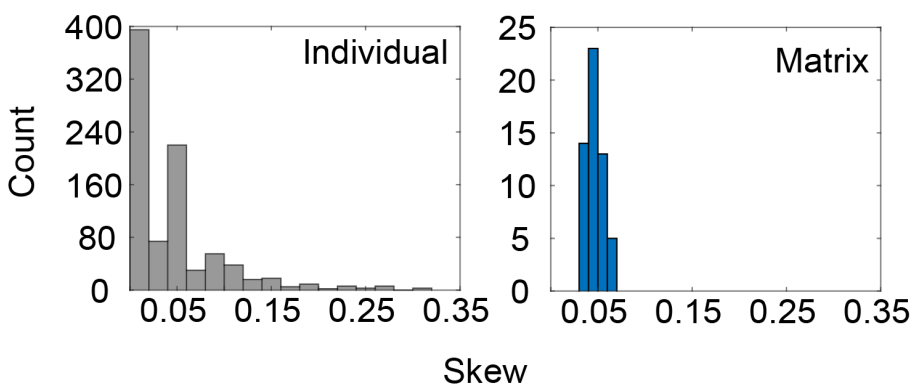

**Figure S4**. **Distribution of individual and aggregate skew**. Left, distribution of skew from each mass and velocity from all combinations of silanes, resulting in 16 counts per given combination of silane. Right, aggregate distribution of skew by summing all sixteen combinations of masses and velocity for a given combination of silanes.

**S6. Power analysis for number of subjects needed in human testing**

We estimated the number of subjects required for an odd-man-out (three-alternative forced choice) psychophysics through a power analysis. Previously, it was shown that humans can distinguish between perflourooctyltricholorsilane and plasma oxidized surfaces.(*10*) Upon inspection, the perflourooctyltrichlorosilane and plasma oxidation generate surfaces with relatively large differences in structures and surface energy. Therefore, we assumed a lower average success rate 60% in our proposed tests, but a similar standard deviation of 15%. Given that chance is 33%, at a power of 0.95, the power analysis indicates that *n*=7 subjects are necessary. However, assumed independence the task as a sequence of five Bernoulli trials, we exceeded the number required in power test by having each subject perform five trials.

*Human psychophysical testing.* Subjects were asked to perform an "odd-man-out" test. In an odd-man-out test, three samples are presented at the same time to a subject. Two of these samples are the same while the third sample has a different surface coating. The location of the different sample and the type of coating was randomized across trials and subjects. Subjects could freely explore the three samples and then asked which sample was different from the other two. Typical exploration time for a comparison took 10-30 seconds.